\documentclass[%
 aip,
 jmp,%
 amsmath,amssymb,
 reprint,%
]{revtex4-1}

\usepackage{graphicx}
\usepackage{dcolumn}
\usepackage{bm}

\usepackage[cal=boondox]{mathalfa}

\DeclareMathAlphabet\mathbfcal{OMS}{cmsy}{b}{n}

\begin{document}

\preprint{AIP/123-QED}

\title[The Hamilton-Jacobi Equation and its Application to ...]{The Hamilton-Jacobi Equation and its Application to Nonlinear Beam Dynamics: Comparison of Approaches}

\author{Stephan I. Tzenov}
\email{tzenov@jinr.ru}
\affiliation{Veksler and Baldin Laboratory for High Energy Physics, Joint Institute for Nuclear Research, 6 Joliot-Curie Street, Dubna, Moscow Region, Russian Federation, 141980}




\date{\today}

\begin{abstract}
The rarely used Hamilton-Jacobi equation has been utilized as an elegant way to find the trajectories of mechanical systems and to derive symplectic maps. Further, the exact solution in kick approximation of Hamilton's equations of motion in interaction representation is written as a generalized one-turn twist map. 
        
One can imagine that the nonlinear kick comes first, followed by the one-period rotation along the machine circumference, or a second alternative in which the one-period rotation occurs before the kick. There is a difference in the result of solving Hamilton's equations between the two cases, which is expressed in obtaining a standard forward twist map in the first case, or alternatively a backward map in the second one. This nontrivial and intuitively unclear peculiarity is usually ignored/overlooked in practically all specialized references on the topic. 
        
Finally, the statistical properties and the behavior of the density distribution of a particle beam in configuration space under the influence of an isolated sextupole have been studied.  
\end{abstract}

\maketitle

\section{\label{sec:intro}Introduction}

Although the study of chaotic motions in nonlinear mechanics has dominated the field in recent times, the study of regular motion and its stability is still a pressing issue in a number of sub-fields of plasma physics, accelerator physics, celestial mechanics, fluid mechanics and others. These important applications include wave–particle interactions \cite{chirik,meiss}, magnetic field structure in magnetic confinement devices \cite{rechest,balescu}, transport and mixing in fluids \cite{morris,weiss}, particle motion in accelerators \cite{berg}, and long-time evolution of the solar system \cite{wisdom}. For instance, increasingly difficult problems arise daily in the design of very large particle accelerators and storage rings. In such machines, particles must be kept in tightly confined orbits for enormous time intervals.

To meet these requirements, accelerator designers rely on single-particle tracking, which involves calculating individual trajectories in external fields for a variety of initial conditions by approximately integrating Hamilton's equations of motion. The specific nature of the problem often requires the use of unusual integration methods, such as "{\it kick approximation}", symplectic integration, symplectic mapping methods, etc. Much effort is put into creating numerical integration procedures valid for large time intervals, but inevitably in coexistence with limitations on accuracy, convergence and computational time. In large accelerators, it is difficult (not to say that very often impossible) to track the orbits of individual particles over sufficiently long time intervals in order to assess their stability. Moreover, one can usually afford to try only a limited number of initial conditions, or in other words, a number of particles orders of magnitude less than those actually contained in the beam itself.

The only scheme known to converge is the superconvergent Kolmogorov-Arnold-Moser (KAM) perturbation theory \cite{arnold,gallavotti}, but it has unfortunately received scant attention as a possible computational tool in both accelerator and plasma physics. Each step of the KAM iteration invokes an approximate solution to the Hamilton-Jacobi equation, usually in lowest order of perturbation. With the exception of an early paper by Robert L. Warnock and Ronald D. Ruth \cite{warnock} and a couple subsequent articles by the same authors, the Hamilton-Jacobi equation method has been practically unused in accelerator physics over the years. To the best of our knowledge, this method has received very scant attention in plasma physics \cite{pfirsch}, as well.

The Hamilton-Jacobi equation provides an elegant framework for solving Lagrangian and Hamiltonian systems by transforming them into a partial differential equation, simplifying problems like finding geodesics and offering a link between classical mechanics and quantum mechanics. Its advantages include a wave-like interpretation and the ability to describe families of solutions related to conserved quantities - the second feature especially valuable when it comes to the description of regular motion. In addition, it offers a powerful approach to solving problems in analytical mechanics by transforming complex dynamics into a single partial differential equation. Among the merits of the Hamilton-Jacobi equation, it is necessary to mention the facilitation in finding optimal canonical transformations that simplify the Hamiltonian system, potentially making the equations of motion trivial in the new canonical variables. Finally, it must be pointed out that the Hamilton-Jacobi equation is not always a simplification. In many cases, the Hamilton-Jacobi approach does not inherently simplify the solution of the original Hamiltonian problem. 

The method of canonical transformations and the associated Hamilton-Jacobi equation are particularly valuable for constructing Hamiltonian maps. It does not suffer from the disadvantages of other known approaches, in which the derivation (except for a known limited number of cases) restricts the possible symplectic forms of the map, thus remaining largely intangible. A good comparison between various methods for deriving symplectic maps, replete with many concrete examples, with an emphasis on the canonical transformations approach, can be found in Ref. \cite{abdul}. 

In the present work, we will show that the above-mentioned method is based on a canonical change of variables, which at first glance eliminates perturbations in periodic time intervals. This procedure transforms the perturbed system into a new one also known as a Hamiltonian system in interaction representation. For that new system the motion is unperturbed (well known in explicit form) throughout the entire period, except for discrete periodic time instants, where all perturbations act instantaneously as kicks. 
 
The article is organized as follows. In Sections \ref{sec:hamjac} and \ref{sec:perhamjac} the Hamilton-Jacobi equation has been introduced as an elegant way to find the trajectories of mechanical systems. Further, on the example of an isolated, infinitely thin magnetic sextupole, the Hamilton-Jacobi equation is being solved perturbatively. The full canonical transformation turns out to be equivalent to the {\it H\'enon map in a canonical form}. In Section \ref{sec:hamhen} the solution of Hamilton's equations of motion in interaction representation has been obtained in the form of a generalized one-turn map. The difference between the case where the nonlinear kick comes first, followed by a single one-period rotation, versus the case where the rotation occurs before the kick has been stressed out. This non-obvious difference is usually tactfully omitted in almost all references devoted to the topic. In Section \ref{sec:statdescr} the statistical properties and the behavior of the density distribution of a particle beam in configuration space under the influence of an isolated sextupole have been studied. Finally, Section \ref{sec:conclude} provides concluding remarks and outlook.


\section{\label{sec:hamjac}The Classical Hamilton-Jacobi Equation}

For simplicity, let us consider a single degree of freedom betatron motion in the horizontal direction of a plane transverse to the particle trajectory in the presence of a singly located sextupole and/or octupole, perturbing the linear accelerator lattice. In what follows, each of the above-mentioned nonlinearities will be considered either separately or in combination in more detail. It is important to note that in a similar way higher-order multipoles can also be included in the consideration, at the cost of increasing algebraic tediousness in the calculations with the order of the multipole. The Hamiltonian governing the single-particle dynamics is set by the expression \cite{tzenovBOOK} 
\begin{equation}
H = {\frac {{\dot{\chi}} {\left( \theta \right)}} {2}} {\left( P^2 + X^2 \right)} + {\frac {{\mathcal{S}}_0 {\left( \theta \right)}} {3}} X^3 + {\frac {{\mathcal{O}}_0 {\left( \theta \right)}} {4}} X^4, \label{Hamilton}
\end{equation} 
where 
\begin{equation}
{\mathcal{S}}_0 {\left( \theta \right)} = {\frac {\lambda_0 {\left( \theta \right)} \beta^{3/2} {\left( \theta \right)}} {2 R^2}}, \qquad {\mathcal{O}}_0 {\left( \theta \right)} = {\frac {\mu_0 {\left( \theta \right)} \beta^2 {\left( \theta \right)}} {6 R^3}}. \label{NotatScal}
\end{equation}
In addition, ${\left( X, P \right)}$ denotes the normalized transverse phase-space coordinates and $\theta$ is the independent azimuthal variable matching the machine circumference, which usually plays the role of time in accelerator theory. Moreover, ${\dot{\chi}} = R / \beta$ is the derivative of the phase advance with respect to the azimuthal variable, where $R$ is the mean machine radius, and $\beta$ is the well-known Twiss beta-function. The dimensionless quantities $\lambda_0 {\left( \theta \right)}$ and $\mu_0 {\left( \theta \right)}$ measure the sextupole and the octupole strengths, respectively, and are given by 
\begin{eqnarray}
\lambda_0 = {\frac {R^2} {B_z}} {\left( {\frac {\partial^2 B_z} {\partial x^2}} \right)}_{x=z=0}, \label{Sextupole} 
\\ 
\mu_0 = {\frac {R^3} {B_z}} {\left( {\frac {\partial^3 B_z} {\partial x^3}} \right)}_{x=z=0}. \label{Octupole}
\end{eqnarray}

The Hamilton-Jacobi equation is an elegant way to find the trajectories of mechanical systems, but unfortunately it is hardly ever used either in the theory of charged particle accelerators, or in the physics of plasmas. As is commonly known \cite{landau,goldstein,bahram}, the basis of the Hamilton-Jacobi method is the introduction of an appropriately chosen generating function that sets the new Hamiltonian to zero. This means that the new coordinates and the new momenta are constants of motion. The Hamilton-Jacobi equation can solve for these constants of motion, providing an alternative to solving differential (Hamilton's or Lagrange's) equations of motion, especially in complex systems. Choosing the generating function to be of the second kind $F {\left( X, p; \theta \right)}$, we can write the Hamilton-Jacobi equation as 
\begin{equation}
\partial_{\theta} F + {\frac {{\dot{\chi}}} {2}} {\left[ {\left( \partial_X F \right)}^2 + X^2 \right]} + {\frac {{\mathcal{S}}_0} {3}} X^3 + {\frac {{\mathcal{O}}_0} {4}} X^4 = 0. \label{HamilJaco}
\end{equation}
Here, $\partial_{u}$ denotes partial derivative with respect to the variable indicated. It is convenient to pass to the phase advance $\chi$ as a new independent variable playing the role of time 
\begin{equation}
\partial_{\chi} F + {\frac {1} {2}} {\left[ {\left( \partial_X F \right)}^2 + X^2 \right]} + {\mathcal{F}} X^3 + {\mathcal{G}} X^4 = 0, \label{HamJacPA}
\end{equation}
where the normalized sextupole and octupole strengths, respectively 
\begin{equation}
{\mathcal{F}} {\left( \theta \right)} = {\frac {{\mathcal{S}}_0 {\left( \theta \right)} \beta {\left( \theta \right)}} {3 R}}, \qquad {\mathcal{G}} {\left( \theta \right)} = {\frac {{\mathcal{O}}_0 {\left( \theta \right)} \beta {\left( \theta \right)}} {4 R}}, \label{SeStrength}
\end{equation}
are periodic in the azimuth ${\mathcal{F}} {\left( \theta + 2 \pi \right)} = {\mathcal{F}} {\left( \theta \right)}$, and ${\mathcal{G}} {\left( \theta + 2 \pi \right)} = {\mathcal{G}} {\left( \theta \right)}$. Finding an exact solution to the above equation represents in itself a rather complex task. The fact that the multipole nonlinearity is usually sufficiently weak (much weaker than the characteristic parameters of the linear magnetic lattice) simplifies this task to some extent and the solution to the Hamilton-Jacobi equation can be sought perturbatively. Since an approximate solution with sufficient accuracy is rather satisfactory in practice, we shall adhere to such a strategy here in the subsequent exposition.

To begin with, considering the sextupole and the octupole as a first-order perturbation, the linear magnetic structure is described by the zero-order generating function
\begin{equation}
F_0 {\left( X, p; \chi \right)} = {\frac {X p} {\cos \chi}} - {\frac {\tan \chi} {2}} {\left( p^2 + X^2 \right)}. \label{LinGenFun}
\end{equation}
The generating function given by the above Eq. \eqref{LinGenFun} is an exact solution to Eq. \eqref{HamJacPA} for ${\mathcal{F}} = {\mathcal{G}} = 0$. The relationship between the old and new canonical variables is 
\begin{equation}
X = x \cos \chi + p \sin \chi, \quad \quad P = - x \sin \chi + p \cos \chi, \label{ZeroSol}
\end{equation}
where as mentioned above the new canonical coordinates ${\left( x, p \right)}$ are constants of motion. This is a well-known result \cite{tzenovBOOK} that represents the motion of particles in an accelerator as a rotation in the normalized phase space by an angle equal to the corresponding phase advance.


\section{\label{sec:perhamjac}Perturbation Solution of the Hamilton-Jacobi Equation}

Let $G {\left( X, p; \chi \right)}$ be the first-order generating function, where $F = F_0 + G + \dots$ and the dots imply higher order contributions. The equation it satisfies is written in the form 
\begin{eqnarray}
\partial_{\chi} G + {\left( {\frac {p} {\cos \chi}} - X \tan \chi \right)} \partial_X G \nonumber 
\\ 
+ {\mathcal{F}} {\left( \theta \right)} X^3 + {\mathcal{G}} {\left( \theta \right)} X^4  = 0. \label{HamJacFO}
\end{eqnarray}
Our goal here is to demonstrate the detailed performance and efficiency of the Hamilton-Jacobi method on the simplest example of lowest-order (cubic) nonlinearity. The solution of the first-order Hamilton-Jacobi equation \eqref{HamJacFO} is sought in the form of a homogeneous polynomial of third order in the mixed canonical variables 
\begin{eqnarray}
G {\left( X, p; \chi \right)} = {\frac {A {\left( \chi \right)}} {3}} X^3 + B {\left( \chi \right)} X^2 p \nonumber 
\\ 
+ C {\left( \chi \right)} X p^2 + {\frac {D {\left( \chi \right)}} {3}} p^3. \label{HamJacFOSol}
\end{eqnarray}
As can be easily verified, the polynomial coefficients satisfy the system of linear first-order differential equations 
\begin{eqnarray}
{\frac {{\rm d} A} {{\rm d} \chi}} - 3 A \tan \chi + 3 {\mathcal{F}} = 0, \label{DiffEqCoeff10}
\\
{\frac {{\rm d} B} {{\rm d} \chi}} - 2 B \tan \chi + {\frac {A} {\cos \chi}} = 0, \label{DiffEqCoeff1}
\end{eqnarray}
\begin{equation}
{\frac {{\rm d} C} {{\rm d} \chi}} - C \tan \chi + {\frac {2 B} {\cos \chi}} = 0, \qquad {\frac {{\rm d} D} {{\rm d} \chi}} + {\frac {3 C} {\cos \chi}} = 0, \label{DiffEqCoeff2}
\end{equation}

In order to illustrate the results obtained by the method of the Hamilton-Jacobi equation, we consider a single sextupole kick (in analogy with Ref. \cite{tzenovDV}) at each successive turn in the vicinity of the locations $\theta = 0, 2 \pi, 4 \pi, \dots$. In thin lens approximation the sextupole strength ${\mathcal{S}}_0 {\left( \theta \right)}$ in Eq. \eqref{NotatScal} can be written as a sampling function (also known as the Dirac comb function) 
\begin{equation}
{\mathcal{S}}_0 {\left( \theta \right)} = {\mathcal{S}} \sum \limits_{k=-\infty}^{\infty} \delta {\left( \theta - 2 k \pi \right)}, \qquad {\mathcal{S}} = {\frac {L_s \lambda_0 \beta_0^{3/2}} {2 R^3}}, \label{DirCombS}
\end{equation}
where $L_s$ is the sextupole length. Taking into account the relation \eqref{NotatScal} and the properties of the Dirac delta function, we can express ${\mathcal{F}} {\left( \chi \right)}$ in terms of the phase advance $\chi$ as an independent variable as follows 
\begin{equation}
{\mathcal{F}} {\left( \chi \right)} = {\frac {\mathcal{S}} {3}} \sum \limits_{k=-\infty}^{\infty} \delta {\left[ \chi {\left( \theta \right)} - k \omega \right]}, \quad \quad \omega = 2 \pi \nu, \label{DirCombF}
\end{equation}
where $\nu$ is the unperturbed betatron tune. The equations \eqref{DiffEqCoeff10} -- \eqref{DiffEqCoeff2} for determining the unknown coefficients $A$, $B$, $C$ and $D$ are coupled linear first-order differential equations. Note that once the solution to a preceding one starting with the first Eq. \eqref{DiffEqCoeff10} for $A$ is found, it automatically determines the solution to the next equation. The general solution to the equation for $A$ can be written as 
\begin{equation}
A {\left( \chi \right)} = - {\frac {3} {\cos^3 \chi}} \int \limits_{}^{\chi} {\rm d} \tau {\mathcal{F}} {\left( \tau \right)} \cos^3 \tau, \label{SolEquA}
\end{equation}
with the natural initial condition $A {\left( 0 \right)} = 0$. To take the above integral in explicit form, we use the well-known Fourier series expansion of the Dirac comb function \cite{tzenovBOOK} 
\begin{equation}
\sum \limits_{k=-\infty}^{\infty} \delta {\left[ \chi {\left( \theta \right)} - k \omega \right]} = {\frac {1} {\omega}} \sum \limits_{n=-\infty}^{\infty} \exp {\left( i n {\frac {\chi} {\nu}} \right)}, \label{ShaFourier}
\end{equation}
as well as the identity 
\begin{eqnarray}
\sum \limits_{k=-\infty}^{\infty} {\frac {{\rm e}^{i k x}} {k + a}} = && \! \! \! {\frac {\pi} {\sin \pi a}} \exp {\left\{ i {\left[ {\left( 2 n + 1 \right)} \pi - x \right]} a \right\}} \nonumber 
\\ 
&& {\rm for} \; 2 \pi n < x < 2 \pi {\left( n + 1 \right)}. \label{Ident}
\end{eqnarray}
The result is 
\begin{eqnarray}
A {\left( \chi \right)} = - {\frac {{\mathcal{M}}_n} {\cos^3 \chi}}, \nonumber 
\\ 
{\mathcal{M}}_n = {\frac {{\mathcal{S}}} {8}} {\left[ {\mathcal{D}}_n {\left( 3 \omega \right)} + 3 {\mathcal{D}}_n {\left( \omega \right)} \pm 4 \right]}, \label{SolCoefA}
\end{eqnarray}
where 
\begin{equation}
{\mathcal{D}}_n {\left( \xi \right)} = \sum \limits_{k=-n}^{n} {\rm e}^{i k \xi} = {\frac {\sin {\left[ {\left( n + 1/2 \right)} \xi \right]}} {\sin {\left( \xi / 2 \right)}}}, \label{Dirichl}
\end{equation}
is the well-known Dirichlet kernel. The interested reader is referred to Ref. \cite{dirichlet}, where the mathematical details concerning the Dirichlet kernel are presented in a good, consistent and understandable form. In addition, the sign "$+$" must be adopted in case the zero phase advance count starts in a small $\epsilon$-neighborhood including the zero ${\left( 0 - \epsilon \right)}$, while the "$-$" sign is taken if the count starts from $0 + \epsilon$ excluding the zeroth kick. Next, the solutions for $B$, $C$ and $D$ are obtained in a straightforward manner and are expressed as follows 
\begin{equation}
B {\left( \chi \right)} = {\frac {{\mathcal{M}}_n} {\cos^2 \chi}} \tan \chi, \qquad C {\left( \chi \right)} = - {\frac {{\mathcal{M}}_n} {\cos \chi}} \tan^2 \chi, \label{SolCoefBC}
\end{equation}
\begin{equation}
D {\left( \chi \right)} = {\mathcal{M}}_n \tan^3 \chi. \label{SolCoefD}
\end{equation}
At this point we need to clarify what the subscript $n$ represents in the definition \eqref{SolCoefA} of the quantity ${\mathcal{M}}_n$. The discrete nature of the sextupole nonlinearity given by the Dirac comb \eqref{DirCombF}, causes discreteness in the solution of the Hamilton-Jacobi equation, so the subscript $n$ denotes the ordinal number of the revolution. 

Let us now explain in detail what the value of the just obtained solution to the Hamilton-Jacobi equation is and what the direct consequence of it is. From Eqs. \eqref{LinGenFun} and \eqref{HamJacFOSol}, we obtain the canonical transformation 
\begin{equation}
x = {\frac {X} {\cos \chi}} - p \tan \chi + B X^2 + 2 C X p + D p^2, \label{CanonTransX}
\end{equation}
\begin{equation}
P = {\frac {p} {\cos \chi}} - X \tan \chi + A X^2 + 2 B X p + C p^2. \label{CanonTransP}
\end{equation}
The first thing that catches the eye is the resonant nature of the canonical transformations \eqref{CanonTransX} and \eqref{CanonTransP} embedded in the quantity ${\mathcal{M}}_n$ entering the corresponding polynomial coefficients. The reason is the specific behavior of the Dirichlet kernel ${\mathcal{D}}_n {\left( 3 \omega \right)}$ for values of the betatron tune close to third-order resonance, and for large values of the number of turns $n$. For small values of the sextupole strength ${\mathcal{S}}$ and for a small number of revolutions, the quantity ${\mathcal{M}}_n$ remains sufficiently small. In such a case, equations \eqref{CanonTransX} and \eqref{CanonTransP} can be solved approximately in explicit form by successive iterations of the zero-order canonical transformation \eqref{ZeroSol}. As a result, we obtain 
\begin{equation}
X \longrightarrow x \cos \chi + {\left( p - {\mathcal{M}}_n x^2 \right)} \sin \chi, \label{CanHenonX}
\end{equation}
\begin{equation}
p - {\mathcal{M}}_n x^2 \longrightarrow  X \sin \chi + P \cos \chi. \label{CanHenonP}
\end{equation}
As we shall see in the next Section, the above canonical transformations are equivalent to the H\'enon map in a canonical form. 


\section{\label{sec:hamhen}Hamilton's Equations of Motion in Interaction Representation and the H\'enon Map}

\subsection{\label{sec:hameq}Hamilton's Equations of Motion}

The generating function \eqref{LinGenFun} eliminates the part in the Hamiltonian \eqref{Hamilton} responsible for the linear structure of the accelerator, thus emphasizing the influence of the relevant nonlinearities. Such a representation is known in classical (and quantum) mechanics as the interaction representation. The new Hamiltonian is written as follows 
\begin{eqnarray}
H_1 {\left( x, p; \theta \right)} = {\frac {{\mathcal{S}}_0 {\left( \theta \right)}} {3}} {\left( x \cos \chi + p \sin \chi \right)}^3 \nonumber 
\\ 
+ {\frac {{\mathcal{O}}_0 {\left( \theta \right)}} {4}} {\left( x \cos \chi + p \sin \chi \right)}^4
. \label{NewHamilt}
\end{eqnarray}
The Hamilton's equations of motion in terms of the new canonical variables ${\left( x, p \right)}$ acquire rather symmetric form 
\begin{eqnarray}
{\frac {{\rm d} x} {{\rm d} \theta}} = {\left[ {\mathcal{S}}_0 {\left( x \cos \chi + p \sin \chi \right)}^2 \right.} \nonumber 
\\ 
{\left. + {\mathcal{O}}_0 {\left( x \cos \chi + p \sin \chi \right)}^3 \right]} \sin \chi, \label{HamEqux}
\end{eqnarray}
\begin{eqnarray}
{\frac {{\rm d} p} {{\rm d} \theta}} = - {\left[ {\mathcal{S}}_0 {\left( x \cos \chi + p \sin \chi \right)}^2 \right.} \nonumber 
\\ 
{\left. + {\mathcal{O}}_0 {\left( x \cos \chi + p \sin \chi \right)}^3 \right]} \cos \chi, \label{HamEqup}
\end{eqnarray}
Let us consider as above a single multipole kick at each successive turn in the vicinity of the locations $\theta = 0, 2 \pi, 4 \pi, \dots$. 

In the subsequent exposition worked out in detail will be the case of an isolated infinitely thin sextupole with a strength ${\mathcal{S}}_0 {\left( \theta \right)}$ as in Eq. \eqref{NotatScal}, written again as a sampling function \eqref{DirCombS}. Similar arguments and considerations are valid in the case of a single octupole kick, or a higher-order-multipole kick. 

\subsection{\label{sec:henonmap}The Standard and the Backward Henon Map}

If the nonlinear magnetic elements can be regarded as concentrated in a point resembling infinitely thin lenses, Hamilton's equations \eqref{HamEqux} and \eqref{HamEqup} can be solved exactly within one revolution. This is a standard procedure where the solution is represented as a recurrent symplectic map. The solutions of the Hamilton's equations can be sought in two alternative intervals $\theta \in {\left( - \epsilon, \; \; \; 2 \pi - \epsilon \right)}$ or $\theta \in {\left( \epsilon, \; \; \; 2 \pi + \epsilon \right)}$ both covering one complete revolution period. In the first case, the lap revolution is counted starting in the epsilon-neighborhood before the nonlinear kick (to be included), while in the second case, the count starts immediately after the previous (uncounted) kick and ends immediately after the corresponding next-in-order kick is performed. In other words, in the first case the nonlinear kick comes first, followed by the one-turn rotation, while in the second case, the one-period rotation occurs before the kick. There is a difference in the result of solving Hamilton's equations between the two cases, which is usually ignored or silenced tactfully in practically all dedicated references on this topic. We will dwell on this difference in more detail here. 

Consider an isolated nonlinear element, say a sextupole - higher-order nonlinear elements can be treated in a completely analogous way. First, we solve Hamilton's equations of motion in the interval $\theta \in {\left( - \epsilon, \; \; \; 2 \pi - \epsilon \right)}$ and obtain 
\begin{equation}
x = x_0 = {\rm const}, \qquad p = p_0 - {\mathcal{S}} x_0^2. \label{SolHamil1}
\end{equation}
Generalizing the above result for each successive $n$-th turn and taking into account the relations of Eqs. \eqref{ZeroSol}, the one-turn map can be written as 
\begin{eqnarray}
X_{n+1} = X_n \cos \omega + {\left( P_n - {\mathcal{S}} X_n^2 \right)} \sin \omega, \nonumber 
\\ 
P_{n+1} = - X_n \sin \omega + {\left( P_n - {\mathcal{S}} X_n^2 \right)} \cos \omega. \label{HenonMap}
\end{eqnarray}
Here, ${\left( X_n, P_n \right)}$ are the initial values ${\left( x_0, p_0 \right)}$ of the canonical variables before the respective $n$-th lap, while ${\left( X_{n+1}, P_{n+1} \right)}$ are the corresponding values after performing the rotation. The two-dimensional mapping \eqref{HenonMap} is known as the H\'enon map \cite{henon,tzenovBOOK}. 
\begin{figure}[ht]
  \includegraphics[width=\linewidth]{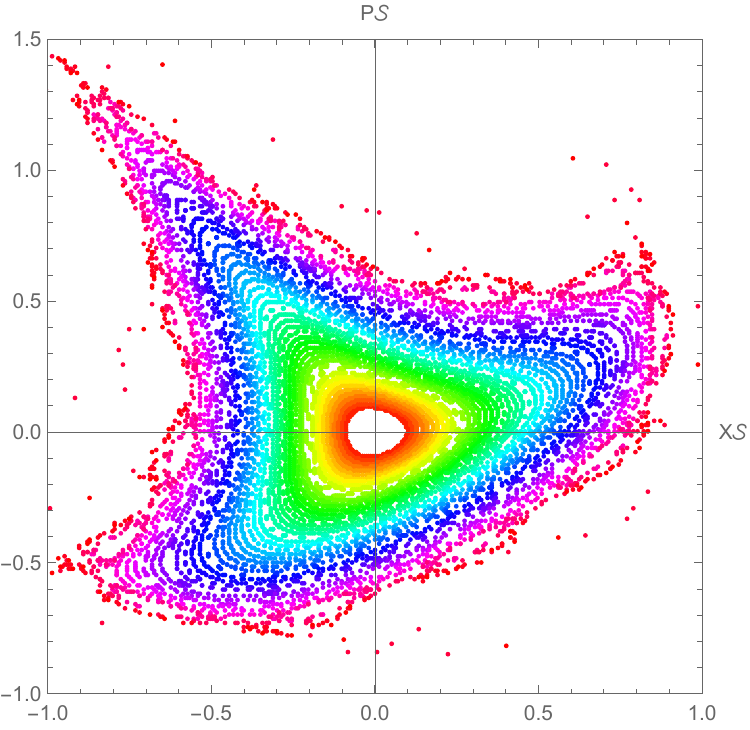}
	\caption{\protect Phase portrait of the canonical H\'enon map \eqref{CanonHenMap} close to the third-order nonlinear resonance $3 \nu = {\rm integer}$. The particular value of the fractional part of the unperturbed betatron tune is taken to be $0.31114$.}
	\label{fig:1} 
\end{figure}
In dimensionless variables ${\left( {\widehat{X}}, {\widehat{P}} \right)} = {\mathcal{S}} {\left( X, P \right)}$ the H\'enon map acquires the canonical form
\begin{eqnarray}
{\widehat{X}}_{n+1} = {\widehat{X}}_n \cos \omega + {\left( {\widehat{P}}_n - {\widehat{X}}_n^2 \right)} \sin \omega, \nonumber 
\\ 
{\widehat{P}}_{n+1} = - {\widehat{X}}_n \sin \omega + {\left( {\widehat{P}}_n - {\widehat{X}}_n^2 \right)} \cos \omega. \label{CanonHenMap}
\end{eqnarray}

Let us now obtain the solution of Hamilton's equations \eqref{HamEqux} and \eqref{HamEqup} in the interval $\theta \in {\left( \epsilon, \; \; \; 2 \pi + \epsilon \right)}$. The result is 
\begin{eqnarray}
x = x_0 + {\mathcal{S}} {\left( x_0 \cos \omega + p_0 \sin \omega \right)}^2 \sin \omega, \nonumber 
\\ 
p = p_0 - {\mathcal{S}} {\left( x_0 \cos \omega + p_0 \sin \omega \right)}^2 \cos \omega. \label{SolHamil2}
\end{eqnarray}
Taking again into account the relations of Eqs. \eqref{ZeroSol}, the one-turn map can be written alternatively  
\begin{eqnarray}
X_{n+1} = X_n \cos \omega + P_n \sin \omega, \nonumber 
\\ 
P_{n+1} = - X_n \sin \omega + P_n \cos \omega - {\mathcal{S}} X_{n+1}^2. \label{RevHenMa}
\end{eqnarray}
This is the {\it backward H\'enon map}, which can be flipped over (reversed) and written like this
\begin{eqnarray}
X_n = X_{n+1} \cos \omega - {\left( P_{n+1} + {\mathcal{S}} X_{n+1}^2 \right)} \sin \omega, \nonumber 
\\ 
P_n = X_{n+1} \sin \omega + {\left( P_{n+1} + {\mathcal{S}} X_{n+1}^2 \right)} \cos \omega. \label{RevHenMap}
\end{eqnarray}
The backward map is basically the same H\'enon map, but with the difference that the motion takes place in the opposite direction, equivalent to reversing time.
\begin{figure}[ht]
  \includegraphics[width=\linewidth]{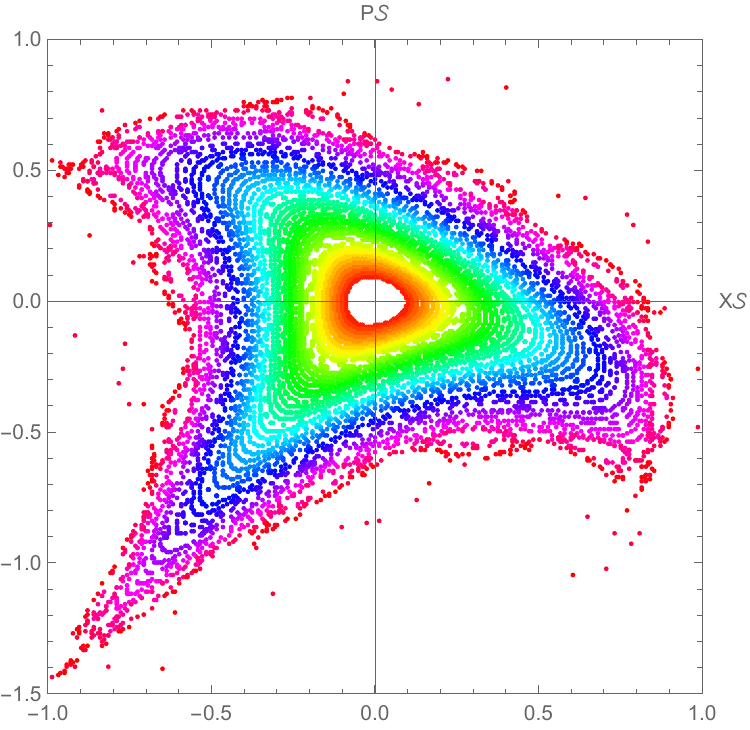}
	\caption{\protect Phase portrait of the backward canonical H\'enon map counterpart of Eqs. \eqref{RevHenMap} close to the third-order resonance $3 \nu = {\rm integer}$. Similar to Fig. \ref{fig:1}, the particular value of the fractional part of the unperturbed betatron tune is taken to be $0.31114$.}
	\label{fig:2} 
\end{figure}
The result just obtained is quite interesting, to some extent non-trivial, and in addition, rather intuitively unexpected. 

Figures (\ref{fig:1}) and (\ref{fig:2}) present the typical phase portraits of the standard H\'enon map \eqref{CanonHenMap} and the backward H\'enon map, respectively. It is clearly visible that apart from a rotation in phase space by an angle of $\pi / 2$, the two phase portraits are identical. 


\section{\label{sec:twistmap}The General Twist Map}

Consider now a combination of nonlinear magnetic elements, say a sextupole (located at $\theta_s = 0$) and an octupole located at $\theta_o$ along the machine circumference. Similar to Eq. \eqref{DirCombS}, in the thin lens approximation the corresponding octupole strength ${\mathcal{O}}_0 {\left( \theta \right)}$ can be written as 
\begin{equation}
{\mathcal{O}}_0 {\left( \theta \right)} = {\mathcal{O}} \sum \limits_{k=-\infty}^{\infty} \delta {\left( \theta - 2 k \pi \right)}, \qquad {\mathcal{O}} = {\frac {L_o \mu_0 \beta_0^2} {6 R^4}}, \label{DirCombOct}
\end{equation}
where $L_o$ is the octupole length. Repeating the arguments used in the derivation of the Henon map, successively for the sextupole and for the octupole, we can write the canonical (dimensionless) one-turn map in the form of a generalized twist map 
\begin{eqnarray}
X_{n+1} = {\left( X_n + F_n \right)} \cos \omega + {\left( P_n + G_n \right)} \sin \omega, \nonumber 
\\ 
P_{n+1} = - {\left( X_n +F_n \right)} \sin \omega + {\left( P_n + G_n \right)} \cos \omega, \label{GenTwistMap}
\end{eqnarray}
where $F_n = F {\left( X_n, P_n \right)}$ and $G_n = G {\left( X_n, P_n \right)}$ are certain functions of $X_n$ and $P_n$. In our case these functions are expressed as 
\begin{eqnarray}
F {\left( X, P \right)} = \Lambda_o {\left[ X \cos \omega_o + {\left( P - X^2 \right)} \sin \omega_o \right]}^3 \nonumber 
\\ 
\times \sin \omega_o, \label{TwistFunc1}
\\ 
G {\left( X, P \right)} = - X^2 - \Lambda_o {\left[ X \cos \omega_o \right.} \nonumber 
\\ 
{\left. + {\left( P - X^2 \right)} \sin \omega_o \right]}^3 \cos \omega_o. \label{TwistFunc2}
\end{eqnarray}
Here $\omega_o$ is the phase advance in the location of the octupole, while $\Lambda_o = {\mathcal{O}} / {\mathcal{S}}^2$ is a dimensionless parameter comparing in terms of order-of-magnitude the strengths of the sextupole and the octupole. It can be easily verified that the generalized twist map \eqref{GenTwistMap} is symplectic if 
\begin{equation}
{\frac {\partial F} {\partial X}} + {\frac {\partial G} {\partial P}} + {\left\{ F, G \right\}} = 0, \label{SymplCond}
\end{equation}
where 
\begin{equation}
{\left\{ F, G \right\}} = {\frac {\partial F} {\partial X}} {\frac {\partial G} {\partial P}} - {\frac {\partial F} {\partial P}} {\frac {\partial G} {\partial X}}, \label{Poiss}
\end{equation}
is the Poisson bracket. 
\begin{figure}[ht]
  \includegraphics[width=\linewidth]{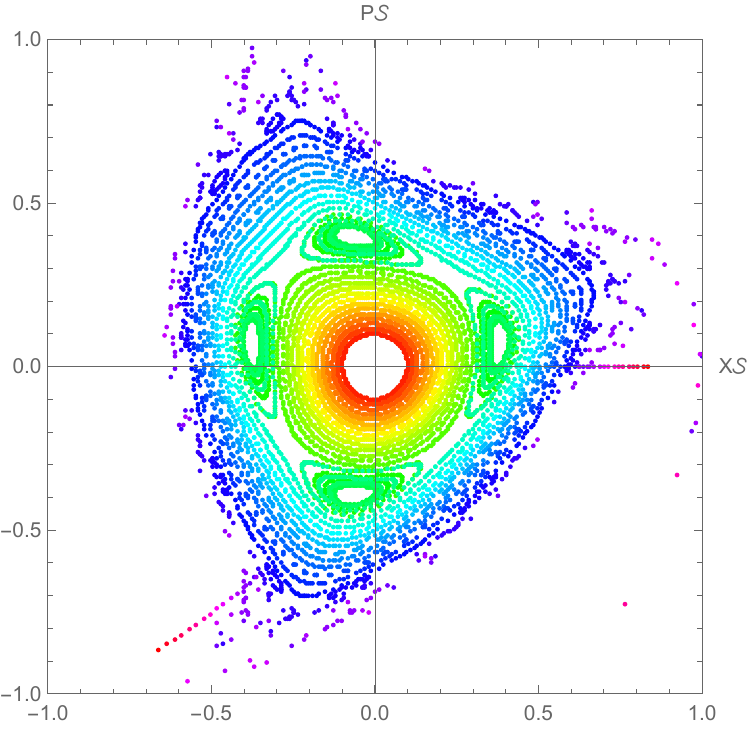}
	\caption{\protect Phase portrait of the generalized twist map of Eqs. \eqref{GenTwistMap} close to the fourth-order resonance $4 \nu = {\rm integer}$. The octupole location is taken to be $\theta_o = \pi$, while its relative strength is $\Lambda_o \sim 1$. The particular value of the fractional part of the unperturbed betatron tune is taken to be $0.243$.}
	\label{fig:3} 
\end{figure}

Figure (\ref{fig:3}) shows the phase portrait of the generalized twist map presented by Eqs. \eqref{GenTwistMap}. The fourfold symmetry of the islands of stability, manifested near the fourth-order resonance driven by the octupole nonlinearity, is clearly visible. The outer envelope (separatrix) of the phase-space curves has a triangular symmetry, caused by the leading sextupole nonlinearity.


\section{\label{sec:statdescr}Statistical Description of Nonlinear Dynamics}

Statistical theory is always associated with a specific dynamical model of some kind, the evolution of which is most often governed by equations of motion. Among the preserved quantities characterizing a given dynamical model, one of particular importance is a quantity possessing classical probabilistic properties, called the distribution function. For Hamiltonian systems of the type \eqref{Hamilton} the distribution function $f {\left( X, P; \theta \right)}$ satisfies the Liouville equation
\begin{equation}
{\frac {\partial f} {\partial \theta}} + {\left\{ f, H \right\}} = 0. \label{Liouville}
\end{equation} 
It can be shown \cite{tzenovBOOK,davidson} that a possible solution of the Liouville equation is a distribution function $f {\left( X, P; \theta \right)}$, which is constant (independent of $X$, $P$ and $\theta$) inside a region in phase space confined by the simply connected boundary curves $P_{(+)} {\left(X; \theta \right)}$ and $P_{(-)} {\left(X; \theta \right)}$, and zero outside. In other words 
\begin{eqnarray}
f {\left( X, P; \theta \right)} = {\mathcal{C}} {\left\{ {\mathcal{H}} {\left[ P - P_{(-)} {\left(X; \theta \right)} \right]} \right.} \nonumber 
\\ 
{\left. - {\mathcal{H}} {\left[ P - P_{(+)} {\left(X; \theta \right)} \right]} \right\}}, \label{WaterBag}
\end{eqnarray} 
where ${\mathcal{H}} {\left( z \right)}$ is the well-known Heaviside function. Defining further the hydrodynamic density $\varrho {\left(X; \theta \right)}$ and the current velocity $V {\left(X; \theta \right)}$ according to the relations 
\begin{equation}
\varrho {\left(X; \theta \right)} = {\mathcal{C}} {\left[ P_{(+)} {\left(X; \theta \right)} - P_{(-)} {\left(X; \theta \right)} \right]}, \label{Density}
\end{equation} 
\begin{equation}
V {\left(X; \theta \right)} = {\frac {1} {2}} {\left[ P_{(+)} {\left(X; \theta \right)} + P_{(-)} {\left(X; \theta \right)} \right]}, \label{CurVel}
\end{equation} 
the Liouville equation \eqref{Liouville} can be cast into completely equivalent system of hydrodynamic equations 
\begin{equation}
{\frac {\partial \varrho} {\partial \chi}} + {\frac {\partial} {\partial X}} {\left( \varrho V \right)} = 0, \label{Contin}
\end{equation} 
\begin{equation}
{\frac {\partial V} {\partial \chi}} + V {\frac {\partial V} {\partial X}} + v_T^2 {\frac {\partial} {\partial X}} {\left( \varrho^2 \right)} = - X - 3 {\mathcal{F}} X^2 - 4 {\mathcal{G}} X^3. \label{CurVelEq}
\end{equation} 
The quantity 
\begin{equation}
v_T^2 = {\frac {1} {8 {\mathcal{C}}^2}}, \label{ThermVel}
\end{equation} 
is the normalized thermal speed-squared. The above system of hydrodynamic equations supplemented with the field equations for the self-consistent potentials has been widely studied in describing collective processes in intense space-charge dominated beams and beam-plasma systems \cite{tzenovBOOK,davidson,tzenvol}. 
\begin{figure}[ht]
  \includegraphics[width=\linewidth]{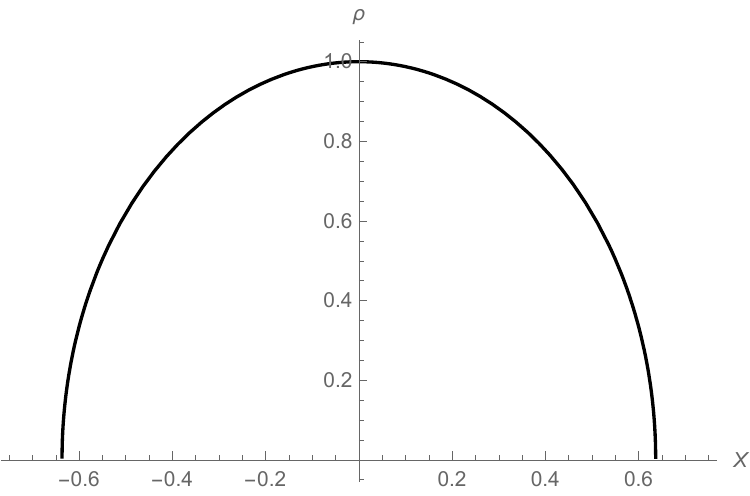}
	\caption{\protect Unperturbed density profile $\varrho_0 {\left( X \right)}$ according to Eq. \eqref{LinearSol}.}
	\label{fig:4} 
\end{figure}

Here it is worth noting in the first place, that in the spirit of Eqs. \eqref{ZeroSol} for the linear unperturbed accelerator lattice, we have
\begin{equation}
P^2 + X^2 = x^2 + p^2 = {\rm const}, \nonumber 
\end{equation} 
implying that $P_{(\pm)} {\left(X; \theta \right)} \sim \pm {\sqrt{{\rm const} - X^2}}$. With this reasoning in hand, it is convenient to write the solution of the hydrodynamic system \eqref{Contin} and \eqref{CurVelEq} in the form
\begin{equation}
\varrho_0^2 {\left( X \right)} = {\mathcal{J}} - {\frac {X^2} {2 v_T^2}}, \qquad \qquad V_0 = 0, \label{LinearSol} 
\end{equation} 
which is exact in the case of an unperturbed linear machine structure. Here ${\mathcal{J}} = {\rm const}$ denotes the linear betatron invariant. This solution will serve us further as a basic zero-order approximation. Considering the nonlinear elements as first-order contributions in magnitude, we write the linearized hydrodynamic equations as 
\begin{equation}
{\frac {\partial \varrho_1} {\partial \chi}} + {\frac {\partial} {\partial X}} {\left( \varrho_0 V_1 \right)} = 0, \label{Contin1}
\end{equation} 
\begin{equation}
{\frac {\partial V_1} {\partial \chi}} + 2 v_T^2 {\frac {\partial} {\partial X}} {\left( \varrho_0 \varrho_1 \right)} = - 3 {\mathcal{F}} X^2 - 4 {\mathcal{G}} X^3. \label{CurVelEq1}
\end{equation} 
Manipulate these in an obvious manner, we end up with a single equation for $\varrho_1$ expressed as 
\begin{eqnarray}
{\frac {\partial^2 \varrho_1} {\partial \chi^2}} - 2 v_T^2 {\frac {\partial} {\partial X}} {\left[ \varrho_0 {\frac {\partial} {\partial X}} {\left( \varrho_0 \varrho_1 \right)} \right]} \nonumber 
\\ 
= {\frac {\partial} {\partial X}} {\left[ \varrho_0 {\left( 3 {\mathcal{F}} X^2 + 4 {\mathcal{G}} X^3 \right)} \right]}. \label{SingRhoEq}
\end{eqnarray} 
It is reasonable to assume that the effect of the nonlinearity on the density function $\varrho_0 {\left( X \right)}$ is limited only to varying the linear invariant $\mathcal{J}$. Thus, we conjecture a tentative solution to the above equation of the form 
\begin{equation}
\varrho_1 {\left( X; \chi \right)} = {\frac {1} {2 \varrho_0}} {\mathcal{J}}_1 {\left( X; \chi \right)}. \label{TentConj}
\end{equation} 
The new unknown function ${\mathcal{J}}_1$ satisfies the following equation 
\begin{eqnarray}
{\frac {\partial^2 {\mathcal{J}}_1} {\partial \chi^2}} - 2 v_T^2 \varrho_0^2 {\frac {\partial^2 {\mathcal{J}}_1} {\partial X^2}} + X {\frac {\partial {\mathcal{J}}_1} {\partial X}} \nonumber 
\\ 
= 2 \varrho_0^2 {\left( 6 {\mathcal{F}} X + 12 {\mathcal{G}} X^2 \right)} \nonumber 
\\ 
- {\frac {1} {v_T^2}} {\left( 3 {\mathcal{F}} X^3 + 4 {\mathcal{G}} X^4 \right)}. \label{TentEqua}
\end{eqnarray} 

Consider again an isolated infinitely thin sextupole; octupoles and other higher-order nonlinear elements can be treated in a similar manner. It is clear that the solution to Eq. \eqref{TentEqua} can be written as follows 
\begin{equation}
{\mathcal{J}}_1 {\left( X; \chi \right)} = {\mathcal{A}} {\left( \chi \right)} X + {\mathcal{B}} {\left( \chi \right)} X^3. \label{TentSolF}
\end{equation} 
where the unknowns ${\mathcal{A}} {\left( \chi \right)}$ and ${\mathcal{B}} {\left( \chi \right)}$ satisfy the equations 
\begin{equation}
{\frac {{\rm d}^2 {\mathcal{B}}} {{\rm d} \chi^2}} + 9 {\mathcal{B}} = - {\frac {9 {\mathcal{F}}} {v_T^2}}, \label{EquatB}
\end{equation} 
\begin{equation}
{\frac {{\rm d}^2 {\mathcal{A}}} {{\rm d} \chi^2}} + {\mathcal{A}} = 12 {\mathcal{J}} {\left( {\mathcal{F}} + v_T^2 {\mathcal{B}} \right)}. \label{EquatA}
\end{equation} 
Utilizing the Fourier series expansion of the Dirac comb function \eqref{ShaFourier} the solution of Eq. \eqref{EquatB} in a Fourier series decomposition is found to be 
\begin{eqnarray}
{\mathcal{B}} {\left( \chi \right)} = {\frac {{\mathcal{S}}} {4 \pi v_T^2}} \sum \limits_{n=-\infty}^{\infty} {\left( {\frac {1} {n - 3 \nu}} - {\frac {1} {n + 3 \nu}} \right)} \nonumber 
\\ 
\times \exp {\left( i n {\frac {\chi} {\nu}} \right)}, \label{SolBFour}
\end{eqnarray}
Taking into account the identity \eqref{Ident}, we can convert the Fourier representation \eqref{SolBFour} into a closed form. The result is 
\begin{equation}
{\mathcal{B}} {\left( \chi \right)} = - {\frac {{\mathcal{S}}} {2 v_T^2 \sin {\left( 3 \omega / 2 \right)}}} \cos 3 {\left[ \chi - {\left( n + 1/2 \right)} \omega \right]}, \label{FinalSolB}
\end{equation}
where $n \omega < \chi < {\left( n+1 \right)} \omega$. In a similar way, one can proceed with the analysis of Eq. \eqref{EquatA}, the solution of which is expressed as follows 
\begin{eqnarray}
{\mathcal{A}} {\left( \chi \right)} = {\frac {2 {\mathcal{J}} {\mathcal{S}}} {\sin {\left( \omega / 2 \right)}}} \cos {\left[ \chi - {\left( n + 1/2 \right)} \omega \right]} \nonumber 
\\ 
+ {\frac {3 {\mathcal{J}} {\mathcal{S}}} {4 \sin {\left( 3 \omega / 2 \right)}}} \cos 3 {\left[ \chi - {\left( n + 1/2 \right)} \omega \right]}. \label{FinalSolA}
\end{eqnarray}
\begin{figure}[ht]
  \includegraphics[width=\linewidth]{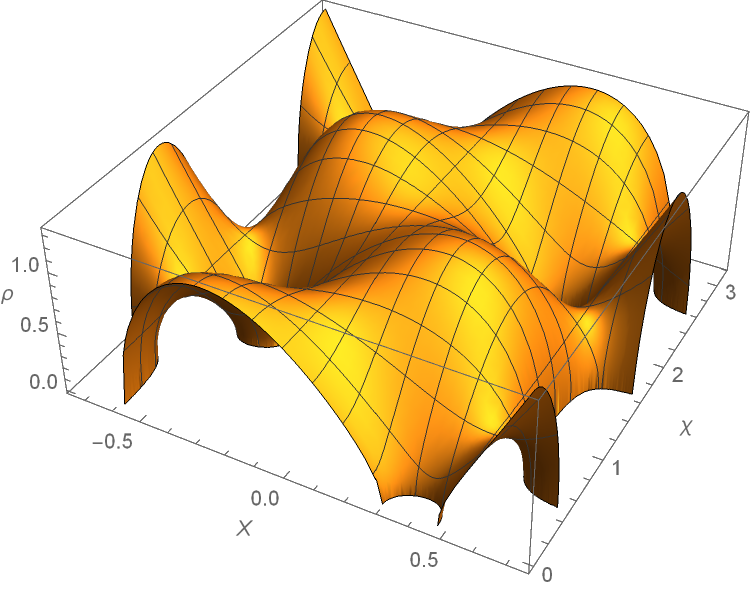}
	\caption{\protect Density profile $\varrho_0 {\left( X, \chi \right)}$, where ${\mathcal{J}}$  has been replaced by ${\mathcal{J}} + {\mathcal{J}}_1$ according to Eq. \eqref{TentSolF}. The particular value of the fractional part of the unperturbed betatron tune is taken to be $0.3333265$, that is close to the third-order resonance.}
	\label{fig:5} 
\end{figure}

Figure (\ref{fig:4}) presents the density function in the linear accelerator structure, while Fig. \ref{fig:5} shows the structural effect of the isolated sextupole. The distortion and the periodicity of the density distribution are clearly visible at values of the unperturbed betatron tune close to third-order resonance. It is important to note that at values of the betatron tune increasingly close to the third-order resonance, unpopulated islands (holes) appear in the density distribution.


\section{\label{sec:conclude}Concluding Remarks and Outlook} 

The main cornerstone of the present article is the introduction of the Hamilton-Jacobi equation as an elegant way to explicitly compute the trajectories of mechanical systems. The application of the canonical transformation approach is particularly advantageous in deriving recurrence maps, since the latter are necessarily symplectic by construction. In particular, on the example of an isolated, infinitely thin magnetic sextupole, the Hamilton-Jacobi equation is solved perturbatively. It turns out that the full canonical transformation thus obtained is equivalent to the H\'enon map in canonical form.

The elimination of the unperturbed part of the overall dynamical system, whose dynamics are assumed to be known, a procedure also known as the interaction representation, is extremely convenient. In addition, if the nonlinear magnetic elements can be considered as infinitely thin lenses, the Hamiltonian equations can be solved exactly within a single one period. Thus, the solution of the equations of motion in the interaction representation is obtained in the form of a generalized one-turn symplectic twist map.

Two cases are considered in detail: in the first case, the full rotation along the machine circumference is counted starting from an epsilon-neighborhood before the nonlinear kick (to be included). In the second case, the counting of one period starts immediately after the previous (uncounted) kick and ends immediately after the corresponding next kick is performed and counted in. The difference in solving Hamilton's equations between the two cases is that the first case yields the classical H\'enon map, while in the second case the backward H\'enon map is obtained. This non-trivial peculiarity is usually ignored or tactfully omitted in practically all specialized references on this topic.

Finally, a study of the statistical properties and behavior of the density distribution of a particle beam in configuration space under the influence of an isolated sextupole is carried out.


\section*{Acknowledgements}
It is a pleasure to express my gratitude to Profs. Jie Gao and Yuan Zhang for making useful comments and suggestions. Fruitful discussions on topics touched upon in the present article with Dr. Yiwei Wang are also gratefully acknowledged.


\bibliographystyle{unsrt}
\bibliography{CLASSICAL_SCATTER}

\end{document}